\documentclass[12pt]{iopart}
\usepackage{amssymb}
\usepackage{iopams}
\usepackage{amsthm}
\usepackage{amscd}
\usepackage{amsfonts}
\usepackage{color}
\usepackage{amsbsy}
\catcode`\@=11
\renewcommand\footnoterule{
 \kern-3\p@
 \hrule\@width.4\columnwidth
 \kern2.6\p@}
\renewcommand\@makefntext[1]{
 \parindent 1em\noindent
 \hb@xt@1.8em{\hss$^{\@thefnmark}$)}\hspace{2pt}
 \footnotesize\rmfamily#1}
\def\@makefnmark{\hspace{.5pt}\hbox{$^{\@thefnmark}$
\hspace{-1pt})}} 
\catcode`\@=11
\renewcommand\footnoterule{
 \kern-3\p@
 \hrule\@width.4\columnwidth
 \kern2.6\p@}
\renewcommand\@makefntext[1]{
 \parindent 1em\noindent
 \hb@xt@1.8em{\hss$^{\@thefnmark}$)}\hspace{2pt}
 \footnotesize\rmfamily#1}
\def\@makefnmark{\hspace{.5pt}\hbox{$^{\@thefnmark}$
\hspace{-1pt}}} 

\newcommand{\cC}{\mathcal{C}}
\newcommand{\cP}{\mathcal{P}}
\newcommand{\cT}{\mathcal{T}}
\newcommand{\cQ}{\mathcal{Q}}
\newcommand{\cS}{\mathcal{S}}
\newcommand{\cL}{\mathcal{L}}
\newcommand{\cD}{\mathcal{D}}
\newcommand{\cM}{\mathcal{M}}

\newcommand{\cPT}{\mathcal{PT}}
\newcommand{\cCPT}{\mathcal{CPT}}

\begin{document}
\title[Unbounded $\cC$-symmetries]{Unbounded $\cC$-symmetries and their
nonuniqueness}

\author{Carl M. Bender${}^a$\ and \ Sergii Kuzhel${}^b$}

\address{${}^a$\ Department of Physics, Kings College London, Strand, London
WC2R 1LS, UK\footnote{Permanent address: Department of Physics, Washington
University, St. Louis, MO 63130, USA.}\\
${}^b$\ AGH University of Science and Technology, 30-059 Krakow, Poland}
\eads{\mailto{cmb@wustl.edu},\ \mailto{kuzhel@mat.agh.edu.pl}}

\begin{abstract}
It is shown that if the $\cC$ operator for a $\cPT$-symmetric Hamiltonian with
simple eigenvalues is not unique, then it is unbounded. The fact that the $\cC$
operator is unbounded is significant because, while there is a formal
equivalence between a $\cPT$-symmetric Hamiltonian and a conventionally
Hermitian Hamiltonian in the sense that the two Hamiltonians are isospectral,
the Hilbert spaces are inequivalent. This is so because the mapping from one
Hilbert space to the other is unbounded. This shows that $\cPT$-symmetric
quantum theories are mathematically distinct from conventional Hermitian quantum
theories.
\end{abstract}

\pacs{00.00, 20.00, 42.10}
\submitto{\JPA}

\section{Introduction}
\label{s1}
The Sturm-Liouville differential-equation eigenvalue problem associated with the
non-Hermitian Hamiltonian
\begin{equation}\label{e1}
H=-\frac{d^2}{dx^2}+x^2(ix)^{\varepsilon}\qquad(0<\varepsilon<2)
\end{equation}
has a positive discrete spectrum \cite{R1}. It was conjectured \cite{R2} that
these spectral properties are a consequence of the invariance of $H$ under the
combination of the space-reflection operator $\cP f(x)=f(-x)$ and the
time-reversal operator $\cT f(x)=f^*(x)$; that is $[H,\cPT]=0$.

The $\cPT$-symmetric Hamiltonian $H$ is not Hermitian\footnote{The terms
`Hermitian operator' and `self-adjoint operator' are equivalent.} in the Hilbert
space $L_2(\mathbb{R})$ whose inner product is
\begin{equation}\label{e2}
(f,g)\equiv\int_{\mathbb{R}}dx[\cT f(x)]g(x)\qquad [f,g\in{L_2({\mathbb{R}})}],
\end{equation}
but $H$ is Hermitian with respect to the $\cPT$ inner product
\begin{equation}\label{e3}
(f,g)_{\cPT}\equiv\int_{\mathbb{R}}dx[\cPT f(x)]g(x)\qquad [f,g\in{L_2({
\mathbb{R}}})],
\end{equation}
where $\cPT f(x)=[f(-x)^*]$. The set of functions $L_2(\mathbb{R})$ endowed with
the $\cPT$ inner product (\ref{e2}) is a Krein space \cite{R3} and $H$ is a
Hermitian operator in the Krein space $L_2(\mathbb{R})$ with the $\cPT$ inner
product $(\cdot,\cdot)_\cPT$. The principal difference between the inner product
$(\cdot,\cdot)$ and the $\cPT$ inner product $(\cdot,\cdot)_\cPT$ is that
$(\cdot,\cdot)_\cPT$ is indefinite; that is, there exist nonzero functions
$f\in{L_2(\mathbb{R})}$ such that $(f,f)_\cPT<0$. Proving that the
$\cPT$-symmetric Hamiltonian $H$ in (\ref{e1}) has a positive real spectrum is
mathematically significant, but $H$ does not have any obvious relevance to
physics until it can be shown that $H$ can serve as a basis for a theory of
quantum mechanics. To do so one must demonstrate that the Hamiltonian $H$ is
Hermitian on a Hilbert space (not a Krein space!) that is endowed with an inner
product whose associated norm is positive definite. Only then can one say that
the theory is unitary and that it has a probabilistic interpretation.

These problems can be overcome for the $\cPT$-symmetric Hamiltonian $H$ by
finding a new (hidden) symmetry represented by a linear operator $\cC$, which
commutes with both the Hamiltonian $H$ and the $\cPT$ operator. In terms of
$\cC$ one must construct a $\cCPT$ inner product
\begin{equation}\label{e4}
(f,g)_\cCPT\equiv\int_{\mathbb{R}}dx[\cCPT f(x)]g(x),
\end{equation}
whose associated norm is positive definite and show that $H$ is Hermitian with
respect to $(\cdot,\cdot)_\cCPT$. When such a $\cC$ operator exists, we say that
the $\cPT$-symmetry of $H$ is unbroken. Constructing the $\cC$ operator is
the key step in showing that the time evolution for the Hamiltonian $H$ is
unitary.

There have been many attempts to calculate the operator $\cC$ \cite{R4,R5} or
the metric operator $\Theta=\cC\cP$ \cite{R6} for the various $\cPT$-symmetric
models of interest. Because of the difficulty of the problem ($\cC$ depends on
the choice of $H$), it is not surprising that the majority of the available
results are approximate, usually expressed as leading terms of perturbation
series. However, these investigations have shown that $\cC$ may be unbounded and
that its choice is nonunique \cite{R7}.

In the present paper we study the phenomena of (possible) nonuniqueness and
unboundedness of $\cC$ for $\cPT$-symmetric Hamiltonians in $L_2(\mathbb{R})$.
To this end we establish in Sec.~\ref{s2} a one-to-one correspondence between
the collection of operators $\cC$ and the collection of all possible $\cPT$
orthogonal pairs of maximal positive and maximal negative subspaces of $L_2(
\mathbb{R})$, where positivity (and negativity) is understood with respect to
the $\cPT$ inner product. This is an underlying mathematical structure that
allows one to explain the property of boundedness/unboundedness of the operator
$\cC$.

Our investigations show that this property is crucial. Indeed, if the $\cC$
operator for $H$ is {\it bounded}, then $H$ is Hermitian on a Hilbert space
that coincides with the {\it same set of functions} $L_2(\mathbb{R})$ but is
endowed with the $\cCPT$ inner product that is equivalent to the initial one
$(\cdot, \cdot)$. Thus, the $\cPT$-symmetric Hamiltonian $H$ can be realized as
Hermitian on the same set of states $L_2(\mathbb{R})$ with the help of the
right choice of the bounded metric operator $\Theta=\cC\cP$. In this case (and
only in this case!) the complete set of eigenfunctions $\{f_n\}$ of $H$ gives
rise to a Riesz basis of $L_2(\mathbb{R})$. It should be emphasized that all
previous papers \cite{R6} devoted to the construction of the metric operator
$\Theta$ have dealt with the case of an operator $\cC$ that is bounded.

The situation is completely different if $\cC$ is {\it unbounded} in $L_2
(\mathbb{R})$ (see Sec.~\ref{s3}). In this case the metric operator $\Theta$ is
not defined on all elements of $L_2(\mathbb{R})$ and the $\cCPT$ inner product
is {\it not} equivalent to the initial one $(\cdot,\cdot)$. This leads to the
Hermitian realization of $H$ in a new Hilbert space $\mathfrak{H}$ that does
not coincide with\footnote{Reference \cite{R8} gives a physical discussion of
this phenomenon.} $L_2(\mathbb{R})$. In fact, the common part of spaces
$\mathfrak{H}$ and $L_2(\mathbb{R})$ contains the linear span $\cD=
\mbox{span}\{f_n\}$ of eigenfunctions of $H$, and the completion of $\cD$
with respect to the nonequivalent inner products $(\cdot,\cdot)$ and
$(\cdot,\cdot)_{\cCPT}$ leads to different Hilbert spaces $L_2(\mathbb{R})$ and
$\mathfrak{H}$, respectively. The set of eigenfunctions $\{f_n\}$ loses the
Riesz-basis property in $L_2(\mathbb{R})$, but it turns out to be an orthogonal
basis in the new space $\mathfrak{H}$. Therefore, in contrast to the case of
bounded operators $\cC$, a $\cPT$-symmetric Hamiltonian $H$ with an unbounded
$\cC$ operator cannot be similar to a Hermitian Hamiltonian in the Hilbert
space $L_2(\mathbb{R})$. The nonuniqueness and unboundedness of the $\cC$
operator is discussed in detail in Secs.~\ref{s3} and \ref{s4}. Sec.~\ref{s5}
contains examples of unbounded $\cC$ operators and a brief
summary is given in Sec.~\ref{s6}.

\section{Preliminaries and basic properties of $\cC$}
\label{s2}

We assume that a closed densely defined linear operator $\cC$ in
$L_2(\mathbb{R})$ obeys the relations
\begin{equation}\label{e5}
\cC^2=I, \qquad [\cC, \cPT]=0.
\end{equation}
Moreover, due to the requirement that the $\cCPT$ inner product (\ref{e4})
determines a positive-definite norm, we additionally assume that ${\cC\cP}$ is a
positive Hermitian operator in $L_2(\mathbb{R})$:
\begin{equation}\label{e6}
\cC\cP>0,\qquad(\cC\cP)^\dag=\cC\cP,
\end{equation}
where $\dag$ means the Dirac adjoint in $L_2(\mathbb{R})$ [the adjoint operator
with respect to (\ref{e2})].

The relations in (\ref{e5}) require an additional explanation in the case where
$\cC$ is unbounded. To be precise, the identity $\cC^2=I$ holds on the domain of
definition $\cD(\cC)$ of $\cC$; that is, $\cC : \cD(\cC)\to\cD(\cC)$ and $\cC^2
f=f$ for all $f\in\cD(\cC)$. Similarly, $[\cC,\cPT]=0$ means that $\cPT:\cD(\cC)
\to\cD(\cC)$ and $\cC\cPT{f}=\cPT\cC{f}$ for all $f\in\cD(\cC)$. If $\cC$ is
bounded, then $\cD(\cC)=L_2(\mathbb{R})$ and the relations $\cC^2=I$ and $[\cC,
\cPT]=0$ should hold on the whole $L_2(\mathbb{R})$.

The conditions (\ref{e5}) and (\ref{e6}) are equivalent to
the following presentation of $\cC$:
\begin{equation}\label{e7}
\cC=e^\cQ\cP,
\end{equation}
where $\cQ$ is a Hermitian operator in $L_2(\mathbb{R})$ that anticommutes with
$\cP$ and $\cT$: \ $\{\cQ,\cP\}=\{\cQ,\cT\}=0$.

Our aim now is to establish another description of the operator $\cC$ in
(\ref{e7}) using the geometric properties of the Krein space $L_2(\mathbb{R})$
with the $\cPT$ inner product (\ref{e3}). To this end, we recall \cite{R3} that
a (closed) subspace $\cL$ of the Hilbert space $L_2(\mathbb{R})$ is called {\it
positive} [{\it uniformly positive}] with respect to the $\cPT$ inner product if
\begin{equation*}
(f,f)_\cPT>0 \qquad [(f,f)_\cPT\geq\alpha(f,f)\quad(\alpha>0)]
\end{equation*}
for all functions $f\in\cL\setminus\{0\}$.

A positive [uniformly positive] subspace $\cL$ is called {\it maximal} if $\cL$
is not a proper subspace of a positive [uniformly positive] subspace in $L_2(
\mathbb{R})$. Negative [uniformly negative] subspaces with respect to the
$\cPT$ inner product and the property of their maximality are similarly defined.

Let $\cL_+$ be a maximal positive subspace of $L_2(\mathbb{R})$. Then its $\cPT$
orthogonal complement
\begin{equation*}
\cL_-=\cL_+^{[\bot]}=\{f\in{L_2(\mathbb{R})}\ :\ (f,g)_\cPT=0,\ \forall{g}\in
\cL_+\}
\end{equation*}
is a maximal negative subspace of $L_2(\mathbb{R})$, and the direct $\cPT$
orthogonal sum\footnote{The brackets $[\dot{+}]$ means orthogonality with
respect to $\cPT$ inner product.}

\begin{equation}\label{e8}
\cS=\cL_+[\dot{+}]\cL_-
\end{equation}
is a {\it dense linear set} in the Hilbert space $L_2(\mathbb{R})$.
The set $\cS$ coincides with $L_2(\mathbb{R})$; that is,
\begin{equation}\label{e9}
L_2(\mathbb{R})=\cL_+[\dot{+}]\cL_-,
\end{equation}
if and only if $\cL_+$ is a {\it maximal uniformly positive subspace} with
respect to the $\cPT$ inner product. In that case the subspace $\cL_-$ is a
maximal uniformly negative subspace.

In the Appendix, we prove the following auxiliary results:
\medskip

\noindent
I. Let $\cC$ be determined by (\ref{e7}), where $\cQ$ is a Hermitian operator in
$L_2(\mathbb{R})$ such that $\{\cQ,\cP\}=\{\cQ,\cT\}=0$. Then the subspaces
\begin{equation}\label{e10}
\cL_+=\frac{1}{2}(I+\cC)\cD(\cC),\qquad\cL_-=\frac{1}{2}(I-\cC)\cD(\cC)
\end{equation}
are $\cPT$ invariant (that is, $\cPT\cL_\pm=\cL_\pm$) and they form a $\cPT$
orthogonal sum (\ref{e8}), where $\cL_+$ and $\cL_-$ are respectively, maximal
positive and maximal negative with respect to the $\cPT$ inner product. The
domain of definition $\cD(\cC)$ is described by (\ref{e8}), and the operator
$\cC$ acts as the identity operator on $\cL_+$ and as the minus identity
operator on $\cL_-$.
\medskip

\noindent
II. Let the subspaces $\cL_\pm$ in (\ref{e8}) be $\cPT$ invariant and let an
operator $\cC$ be defined on (\ref{e8}) as mentioned above; that is, $\cD(\cC)=
\cS$ and the restriction of $\cC$ onto $\cL_+$ [$\cL_-$] coincides with the
identity operator [minus identity operator]. Then the operator $\cC$ can also be
determined by (\ref{e7}), where $\{\cQ,\cP\}=\{\cQ,\cT\}=0$.
\medskip

It follows from statements I and II that \emph{there exists a one-to-one
correspondence between the set of operators $\cC=e^\cQ\cP$ with $\{\cQ,\cP\}=\{
\cQ,\cT\}=0$ and the set of $\cPT$ orthogonal decompositions (\ref{e8}), where
$\cL_\pm$ are $\cPT$ invariant and $\cL_+[\cL_-]$ belongs to the collection of
all maximal positive [maximal negative] subspaces with respect to the $\cPT$
inner product.} The action of $\cC=e^\cQ\cP$ is reduced to the $\pm$ identity
operator on $\cL_\pm$.

This relationship allows one to describe various classes of $\cC$. In
particular, \emph{$\cC$ is a bounded operator in $L_2(\mathbb{R})$ if and only
if the corresponding maximal subspace $\cL_+[\cL_-]$ in (\ref{e10}) is uniformly
positive [negative]} \cite{R9}. In this case $\cC$ is determined on the whole
space $L_2(\mathbb{R})$ due to (\ref{e9}).

\section{$\cPT$-symmetric Hamiltonians with $\cC$ operators}
\label{s3}

We begin with some definitions to avoid possible misunderstanding of the results
below.

1. A densely defined operator $H$ in $L_2(\mathbb{R})$ is called a
\emph{$\cPT$-symmetric Hamiltonian} if $[H,\cPT]=0$ and $H$ is Hermitian with
respect to the $\cPT$ inner product (\ref{e3}). The relation $[H,\cPT]=0$ means
that $\cPT : \cD(H)\to\cD(H)$ and $H\cPT{f}=\cPT{H}f$ for all $f\in\cD(H)$.

2. Let $H$ be a $\cPT$-symmetric Hamiltonian and let $\cD$ be a linear subset of
$\cD(H)$ such that the closure of the restriction of $H$ onto $\cD$ coincides
with $H$ [which is the closure of  $H'=H\upharpoonright_{\cD}$ is $H$].

\emph{We say that a $\cPT$-symmetric Hamiltonian $H$ has an operator $\cC=e^\cQ
\cP$ if the commutation relation $H\cC{f}=\cC{H}f$  holds for all $f\in\cD
\subset\cD(H)$.} The latter means that
\begin{equation}\label{e10b}
\cD(\cC)\supset\cD, \quad \cC : \cD\to\cD, \quad H:\cD\to\cD(\cC).
\end{equation}

If $\cC$ is a bounded operator, then the first and third relations in
(\ref{e10b}) are trivial because $\cD(\cC)=L_2(\mathbb{R})$. Moreover, the
commutation relation $H\cC{f}=\cC{H}f$ can be extended onto $\cD(H)$, which
implies that $\cC : \cD(H)\to\cD(H)$. Therefore, in the case of a bounded
operator $\cC$, we can suppose that $\cD=\cD(H)$.

The necessary condition of the existence of a bounded operator $\cC$ for a
$\cPT$-symmetric Hamiltonian $H$ is the following resolvent estimate:
$$\|(H-\lambda{I})^{-1}\|\leq\frac{M}{\textsf{Im} \ \lambda},\qquad\lambda\in
\mathbb{C}_+,$$
where the constant $M>0$ does not depend on the choice of $\lambda$ from the
complex upper-half plane $\mathbb{C}_+$.

The general criterion follows from the results of \cite{R9} and \cite{R9a}.
To be precise, \emph{A $\cPT$-symmetric Hamiltonian $H$ acting in $L_2(
\mathbb{R})$ has a bounded operator $\cC=e^\cQ\cP$ if and only if the spectrum
of $H$ is real and there exists a constant $M$ such that
$$\mathrm{sup}_{\varepsilon>0}\varepsilon\int_{-\infty}^{\infty}\|(H-\lambda{I}
)^{-1}f\|^2d\xi\leq{M}\|f\|^2,\qquad\lambda=\xi+i\varepsilon,\qquad{f}\in{L_2(
\mathbb{R})},$$
where the integral is taken along the line $\lambda=\xi+i\varepsilon$
($\varepsilon>0$ is fixed).}

In the following we assume that a $\cPT$-symmetric Hamiltonian $H$ has a {\it
complete set of eigenfunctions} $\{f_n\}$ in $L_2(\mathbb{R})$. In this context
{\it complete set} means that the linear span of eigenfunctions $\{f_n\}$, that
is, the set of all possible {\it finite} linear combinations
\begin{equation*}
\mbox{span}\{f_n\}=\left\{\sum_{n=1}^d{c_nf_n}\ : \ \forall{d}\in
\mathbb{N},\ \forall{c_n}\in{\mathbb{C}}\right\},
\end{equation*}
is a dense subset in $L_2(\mathbb{R})$.

In general, the completeness of a linearly independent sequence of
eigenfunctions $\{f_n\}$ does not mean that $\{f_n\}$ is a Schauder
basis\footnote{A sequence $\{f_n\}$ is a Schauder basis for $L_2(\mathbb{R})$ if
for each $f\in{L_2(\mathbb{R})}$, there exist unique scalar coefficients $\{c_n
\}$ such that $f=\sum_{n=1}^\infty{c_nf_n}$ \cite{R10}.} of $L_2(\mathbb{R})$.
The difference is that the completeness of $\{f_n\}$ allows us to approximate an
arbitrary $f\in{L_2(\mathbb{R})}$ by finite linear combinations $\sum_{n=1}^d{
c_n^df_n}{\to}f$ as $d\to\infty$, where $c_n^d$ {\it depend on the choice of}
$d$, while the definition of a Schauder basis requires that $c_n^d$ does not
depend on $d$.

Let $H$ be a $\cPT$-symmetric Hamiltonian with complete set of eigenfunctions
$\{f_n\}$ corresponding to real eigenvalues $\{\lambda_n\}$. For simplicity, we
assume that the spectrum of $H$ coincides with the set of eigenvalues
$\{\lambda_n\}$ and $\lambda_n$ are {\it simple} eigenvalues; that is,
$\dim\ker(H-\lambda_n{I})=1$.

Let $\cD$ denote the linear span of all eigenfunctions of $H$. The set $\cD$ is
dense in $L_2(\mathbb{R})$ and the closure of $H'=H\upharpoonright_{\cD}$
coincides with $H$.

Since $H$ is Hermitian with respect to the $\mathcal{PT}$ inner product $(\cdot,
\cdot)_{\mathcal{PT}}$, the eigenfunctions $f_n$ are $\mathcal{PT}$-orthogonal:
$(f_n,f_m)_{\mathcal{PT}}=0$ for $m\not={n}$. Furthermore, every eigenfunction
$f_n$ is either positive $(f_n,f_n)_{\mathcal{PT}}>0$ or negative $(f_n,f_n)_{
\mathcal{PT}}<0$ with respect to $(\cdot, \cdot)_{\mathcal{PT}}$. Indeed, if we
suppose that $(f_n,f_n)_{\mathcal{PT}}=0$, then the eigenfunction $f_n$ is
${\mathcal{PT}}$ orthogonal to $\cD$. Therefore the vector ${\mathcal{P}}f_n$
should be orthogonal to $\cD$ in the sense of initial inner product of $L_2(
\mathbb{R})$. This implies that ${\mathcal{P}}f_n=0$ and hence, $f_n=0$. This
contradicts the assumption that $f_n$ is an eigenfunction.

Let us separate the sequence $\{f_n\}$ by the sign of the $\cPT$ inner products
$(f_n,f_n)_\cPT$:
$$f_{n}=\left\{\begin{array}{l}f_{n}^+ \quad\mbox{if}\quad(f_n,f_n)_\cPT>0, \\
f_{n}^- \quad \mbox{if} \quad (f_{n},f_{n})_\cPT<0\end{array}\right.$$
and denote by  $\cL_+'$ and $\cL_-'$ the closure of $\mbox{span}\{f_n^+\}$ and
$\mbox{span}\{f_n^-\}$ in $L_2(\mathbb{R})$. The $\cPT$ invariant subspaces
$\cL_\pm'$ are positive/negative with respect to $\cPT$ inner product and the
direct $\cPT$ orthogonal sum
\begin{equation}\label{e12}
\cS'=\cL_+'[\dot{+}]\cL_-'
\end{equation}
is a dense set in $L_2(\mathbb{R})$.

If the eigenfunctions $\{f_n\}$ of $H$ turn out to be the Riesz
basis\footnote{A Schauder basis $\{f_n\}$ is a Riesz basis if there exist an
invertible operator $A$ and an orthonormal basis $\{\psi_n\}$ in ${L_2(\mathbb{
R})}$ such that $f_n=A\psi_n$.} of $L_2(\mathbb{R})$, then $\cS'={L_2(\mathbb{R}
)}$ and the subspaces $\cL_\pm'$ are \emph{maximal} uniformly positive/negative
\cite[Chap. 1]{R3}. In that case the decomposition (\ref{e12}) is transformed to
the decomposition (\ref{e9}) and it defines a \emph{bounded} operator $\cC$ in
$L_2(\mathbb{R})$. This operator is a $\cC$ operator for $H$ because the
operator identity $H\cC{f}=\cC{H}f$ and relations (\ref{e10b}) are true for all
elements $f\in\cD$.

If the eigenfunctions $\{f_n\}$ of $H$ do not form the Riesz basis but they are
a Schauder basis, then the subspaces $\cL_\pm'$ lose the property of
\emph{uniform} positivity/negativity but these subspaces are still
\emph{maximal} positive/negative \cite[Chap. 1]{R3}. In this case, the
decomposition (\ref{e12}) takes the form (\ref{e8}). This means that (\ref{e12})
correctly defines a unique unbounded operator $\cC$ for $H$.

In two cases above [Riesz and Schauder bases] the action of $\cC$ is completely
determined by eigenfunctions $\{f_n\}$ of $H$.

In the general case when $\{f_n\}$ is a complete set of eigenvalues, it may
happen that the subspaces $\cL_\pm'$ are \emph{only} positive/negative. Then the
decomposition (\ref{e12}) cannot properly define an operator $\cC$ with
properties (\ref{e5}), (\ref{e6}). To this end we have to extend $\cL_\pm'$ to
maximal positive/negative subspaces. The (possible) nonuniqueness of such kinds
of extensions leads to the nonuniqueness of unbounded operators $\cC$ for $H$.

\section{Reasons for nonuniqueness of ${\cC}$ operators}
\label{s4}

For a given $\cPT$-symmetric Hamiltonian $H$ in $L_2(\mathbb{R})$ there may
exist different $\cC$ operators. Due to statements I and II, the nonuniqueness
of $\cC$ is equivalent to the existence of {\it different} decompositions
(\ref{e8}) [or (\ref{e9}) for the case of bounded $\cC$] that reduce the
operator $H$.

There are \emph{two reasons for the nonuniqueness of the $\cC$ operator} for a
$\cPT$-symmetric Hamiltonian $H$ with a complete set of eigenfunctions $\{f_n\}$
in $L_2(\mathbb{R})$. \emph{One of them can be illustrated even for the matrix
case}, and it deals with the (possible) appearance of nontrivial neutral
elements with respect to the $\cPT$ inner product in at least one of
eigensubspaces $\ker(H-\lambda{I})$.

Let us illustrate this phenomenon by considering a $\cPT$-symmetric Hamiltonian
$H$ with a Riesz basis $\{f_n\}$ of eigenfunctions and hence with a bounded
$\cC$ operator. We assume that the first $k$ eigenfunctions $f_1,\ldots,f_k$
correspond to the eigenvalue $\lambda_0$. For the sake of simplicity, other
eigenvalues $\lambda_1,\lambda_2,\ldots$ of $H$ are assumed to be {\it simple}.
This means that $\ker(H-\lambda_m{I}) \ \ m\in\mathbb{N}$ coincides with the
span of the function $f_{k+m}$.

The bounded operator $\cC$ generates the decomposition (\ref{e9}) of $L_2(
\mathbb{R})$, where subspaces $\cL_\pm$ are determined by (\ref{e10}). Every
eigenfunction $f_n$ is decomposed along (\ref{e9}) as\footnote{One of
functions $f_n^\pm$ may vanish.}
\begin{equation}\label{e15}
f_n=f_n^++f_n^-,\qquad f_n^{\pm}=\frac{1}{2}(I\pm\cC)L_2(\mathbb{R}),
\end{equation}
where $f_n^\pm\in\cL_\pm$. The sequences of functions $\{f_n^\pm\}$ are Riesz
bases of $\cL_\pm$; that is, $\cL_+$ and $\cL_-$ coincide with the closures of
the linear spans of $\{f_n^+\}$ and $\{f_n^-\}$, respectively.

The functions $f_n^\pm$ in (\ref{e15}) are also eigenfunctions of $H$
corresponding to the same eigenvalue. Therefore, due to the {\it simplicity} of
the eigenvalues $\lambda_m$ $(m\in\mathbb{N})$, the decomposition (\ref{e15}) of
the corresponding eigenfunctions $f_{k+m}$ may contain only one nonzero element.
This means that
\begin{equation}\label{e16}
f_{k+m}=\left\{\begin{array}{l} f_{k+m}^+ \quad \mbox{if} \quad (f_{k+m},f_{k+m}
)_\cPT>0, \\ f_{k+m}^- \quad \mbox{if} \quad (f_{k+m},f_{k+m})_\cPT<0
\end{array}\right. \qquad \forall{m}\in\mathbb{N}.
\end{equation}
[The case $(f_{k+m},f_{k+m})_\cPT=0$ is impossible because it gives two linearly
independent eigenfunctions $f_{k+m}^\pm$ of $H$, which contradicts the
simplicity of $\lambda_m$.] Therefore, the functions $f_n^\pm$ are uniquely
determined by $f_n$ when $n=k+m$.

The span of the first $k$ eigenfunctions $f_1,\ldots,f_k$ coincides with $\ker(
H-\lambda_0{I})$. If this finite-dimensional subspace contains nontrivial
neutral elements with respect to the $\cPT$ inner product, then $\ker(H-
\lambda_0{I})$ contains positive elements with respect to the $\cPT$ inner
product as well as negative ones. This means that $\ker(H-\lambda_0)$ admits
{\it different} $\cPT$ orthogonal decompositions onto positive and
negative $\cPT$ invariant subspaces $\cM_\pm$:
\begin{equation*}
\ker(H-\lambda_0)=\cM_+[\dot{+}]\cM_-.
\end{equation*}
One of possible decompositions is $\cM_+=\mbox{span}\{f_n^+\}_{n=1}^k$ and $\cM_
-=\mbox{span}\{f_n^-\}_{n=1}^k$, where the elements $f_n^\pm$ are determined by
$f_n$ with the use of the decomposition (\ref{e15}). Fixing another
decomposition $\ker(H-\lambda_0)=\cM'_+[\dot{+}]\cM'_-$ with $\cPT$ invariant
subspaces $\cM'_\pm$, we obtain other decompositions of the functions
\begin{equation}\label{e17}
f_n={f'}_n^++{f'}_n^-\qquad(n=1,\ldots,k)
\end{equation}
onto positive and negative parts with respect to the $\cPT$ inner product.

Let us define $\cL_+'$ and $\cL_-'$, respectively, as the closure [with respect
to $(\cdot,\cdot)$] of the linear spans of the $\cPT$ orthogonal functions
\begin{equation*}
[\{{f'}_n^+\}_{n=1}^{k}\cup\{f_{k+m}^+\}_{m=1}^\infty]\qquad\mbox{and}\qquad
[\{{f'}_n^-\}_{n=1}^{k}\cup\{f_{k+m}^-\}_{m=1}^\infty].
\end{equation*}
By this construction, $\cL_\pm'$ are maximal positive/negative subspaces with
respect to the $\cPT$ inner product, $\cL_\pm'$ are $\cPT$ invariant, and
\begin{equation}\label{e18}
L_2(\mathbb{R})=\cL_+'[\dot{+}]\cL_-'.
\end{equation}
It is clear that $\cL_+\not=\cL_+'$ and $\cL_-\not=\cL_-'$. Therefore, the
$\cC'$ operator of $H$ determined by (\ref{e18}) [that is, $\cC'
\upharpoonright_{\cL_+'}=I$ and $\cC'\upharpoonright_{\cL_-'}=-I$] does not
coincide with the initial operator $\cC$.

\emph{Another reason leading to the nonuniqueness of $\cC$ cannot be observed
for bounded $\cC$ operators and this phenomenon may appear only for unbounded
operators $\cC$}.

Indeed, let $H$ be a $\cPT$-symmetric Hamiltonian with a complete set of
eigenfunctions $\{f_n\}$ that correspond to the simple eigenvalues of $H$.
First, suppose that $H$ has a bounded $\cC$ operator. Then the decomposition
(\ref{e9}) holds, where $\cL_\pm$ are determined by (\ref{e10}). Doing the
$\cPT$ arrangement of $\{f_n\}$ according to (\ref{e15}), we obtain two
sequences of functions $\{f_n^+\}$ and $\{f_n^-\}$ belonging to $\cL_+$ and
$\cL_-$, respectively. Due to the simplicity of the eigenvalues of $H$, one of
the functions $f_n^+$ and $f_n^-$ in (\ref{e15}) coincides with $f_n$, while
another one is the zero function [see (\ref{e16})]. Thus, the sequences
$\{f_n^{\pm}\}$ are the result of the separation of the sequence $\{f_n\}$ by
the signs of the $\cPT$ inner products $(f_n,f_n)_\cPT$.

For the case of bounded $\cC$, it was shown above that $\{f_n\}$ can be
considered as an orthonormal basis\footnote{After the normalization procedure.}
of $L_2(\mathbb{R})$ with respect to the $\cCPT$ inner product $(\cdot,\cdot
)_\cCPT$. Therefore, the sequences $\{f_n^+\}$ and $\{f_n^-\}$ are orthonormal
bases of the maximal positive subspace $\cL_+$ and the maximal negative subspace
$\cL_-$, respectively. This means that the initial sequence of eigenfunctions
$\{f_n\}$ determines the unique decomposition (\ref{e9}) that leads to the
uniqueness of a bounded operator $\cC$.

To summarize, \emph{for a $\cPT$-symmetric Hamiltonian $H$ with a complete set
of eigenfunctions $\{f_n\}$ corresponding to simple eigenvalues, there may exist
\emph{only one bounded} operator $\cC$}.

However, the situation is completely different for the case of unbounded
operators $\cC$. Precisely, we are going to show below that for a
$\cPT$-symmetric Hamiltonian $H$ with a complete set of eigenfunctions $\{f_n\}$
corresponding to simple eigenvalues, \emph{there may exist infinitely many
unbounded operators $\cC$}. This problem was inspirited by the results of
\cite{R5}, where for the $\cPT$-symmetric Hamiltonian
\begin{equation*}
H=\frac{1}{2}p^2+\frac{1}{2}\mu^2q^2+i\epsilon{q}^3
\end{equation*}
infinitely many operators $\cC$ were constructed by formal perturbative
calculations.

Let us briefly illustrate the principal idea.\footnote{See also the end of
Sec.~3.} Assume that a $\cPT$-symmetric Hamiltonian $H$ with a complete set
of eigenfunctions $\{f_n\}$ corresponding to simple eigenvalues has an unbounded
$\cC$ operator with the set $\cD=\mbox{span}\{f_n\}$ in (\ref{e10b}).

Taking into account that the direct sum (\ref{e8}) of the subspaces $\cL_\pm$
from (\ref{e10}) determines the domain of $\cC$ and repeating the previous
arguments, we separate the sequence $\{f_n\}$ by the sign of the $\cPT$ inner
products $(f_n,f_n)_\cPT$. The obtained sequences $\{f_n^\pm\}$ belong to
$\cL_\pm$.

Let $\cL_+'$ and $\cL_-'$ be the closure of $\mbox{span}\{f_n^+\}$ and $\mbox{
span}\{f_n^-\}$ with respect to the initial inner product $(\cdot,\cdot)$. By
this construction, $\cL_\pm'\subset\cL_\pm$ and the direct $\cPT$ orthogonal sum
\begin{equation}\label{e19}
\cS'=\cL_+'[\dot{+}]\cL_-'
\end{equation}
is a dense set in $L_2(\mathbb{R})$ (due to the completeness of $\{f_n\}$).
However, it may happen that $\cL_\pm'$ are {\it proper subspaces} of $\cL_\pm$
[that is, $\cL_\pm'\subset\cL_\pm$ and $\cL_\pm'\not=\cL_\pm$]. This phenomenon
was first observed by Langer \cite{R12}. His paper provides a mathematically
rigorous explanation based on the fact that the $\cCPT$ inner product $(\cdot,
\cdot)_\cCPT$ is {\it singular} with respect to the initial inner product
$(\cdot,\cdot)$.

If $\cL_\pm'\subset\cL_\pm$, then the positive $\cL_+'$ and negative $\cL_-'$
subspaces with respect to the $\cPT$ inner product do not have the property of
maximality, and hence the direct sum (\ref{e19}) {\it does not define} an
operator $\cC$ with properties (\ref{e5}) and (\ref{e6}). The positive $\cL_+'$
and negative $\cL_-'$ subspaces in (\ref{e19}) can be extended to maximal
positive and maximal negative subspaces in different ways that lead to the
nonuniqueness of $\cC$. (One of the possible extensions are the subspaces
$\cL_\pm$ mentioned above.) These phenomena are discussed in detail in the next
section.

\section{An example of $\cPT$-symmetric Hamiltonians $H$ with different
unbounded operators $\cC$.}
\label{s5}

Let $\{\gamma_n^+\}$ and $\{\gamma_n^-\}$ be orthonormal bases of real functions
in the sets of even $L_2^{even}$ and odd $L_2^{odd}$ functions of $L_2(\mathbb{
R})$, respectively. In particular, we can choose the even and odd Hermite
functions. Any function $\phi\in{L_2(\mathbb{R})}$ has the representation $\phi=
\sum_{n=1}^{\infty}(c_n^+\gamma_n^++c_n^-\gamma_n^-)$, where the sequences
$\{c_n^\pm\}$ are elements of the Hilbert space $l_2$; that is, $\sum_{n=1
}^\infty|c_n^\pm|^2<\infty$. The operator
\begin{equation}\label{e20}
T\phi=\sum_{n=1}^{\infty}i\alpha_n(c_n^+\gamma_n^--c_n^-\gamma_n^+),\qquad
\alpha_n=(-1)^n\left(1-\frac{1}{n}\right)
\end{equation}
plays a key role in our construction and has many useful properties that can be
directly deduced from (\ref{e20}). In particular, $T$ is a Hermitian
contraction in $L_2(\mathbb{R})$ that anticommutes with $\cP$ and $\cT$ (that
is, $H^\dag=H$, $\|T\phi\|<\|\phi\|$ $[\phi\not=0]$, and $\{T,\cP\}=\{T,\cT\}=
0$). The anticommutation with $\cP$ means that $T$ interchanges the sets of
even and odd functions: $T : L_2^{even}{\to}L_2^{odd}$ and vice versa. Denote
\begin{equation}\fl\label{e21}
\cL_+=\{f^+=\gamma^++T\gamma^+:\gamma^+\in{L_2^{even}}\},\quad\cL_-=\{f^-=
\gamma^-+T\gamma^-:\gamma^-\in{L_2^{odd}}\}.
\end{equation}
Since $\{T,\cP\}=\{T,\cT\}=0$, the subspaces $\cL_\pm$ are $\cPT$ invariant and
$${\cPT}f^+={\cT}(\gamma^+-T\gamma^+)\quad [\gamma^+\in{L_2^{even}}],\quad
{\cPT}f^-={\cT}(-\gamma^-+T\gamma^-) \quad [\gamma^-\in{L_2^{odd}}].$$
Hence,
\begin{eqnarray*}
(f^+,f^-)_{\cP\cT}&=&(\gamma^+-T\gamma^+,\gamma^-+T\gamma^-)=(\gamma^+,T
\gamma^-)-(T\gamma^+,\gamma^-)\\ &=&(T\gamma^+,\gamma^-)-(T\gamma^+,\gamma^-)=0.
\end{eqnarray*}
Thus the subspaces $\cL_\pm$ are $\cP\cT$ orthogonal.

The subspace $\cL_+$ is positive with respect to the $\cP\cT$ inner product
$(\cdot,\cdot)_{\cP\cT}$ because
$$(f^+,f^+)_{\cP\cT}=(\gamma^+-T\gamma^+, \gamma^++T\gamma^+)=(\gamma^+,
\gamma^+)-(T\gamma^+, T\gamma^+)=\sum_{n=1}^{\infty}(1-\alpha^2_n)|c_n^+|^2>0,$$
where $\gamma^+=\sum_{n=1}^{\infty}c_n^+\gamma_n^+$. However, $(f^+,f^+)_{\cP
\cT}$ is not topologically equivalent to the initial inner product
$$(f^+,f^+)=(\gamma^++T\gamma^+, \gamma^++T\gamma^+)=(\gamma^+, \gamma^+)+(T
\gamma^+, T\gamma^+)=\sum_{n=1}^{\infty}(1+\alpha^2_n)|c_n^+|^2$$
on $\cL_+$ because $\lim_{n\to\infty}(1-\alpha_n^2)=\lim_{n\to\infty}\frac{1}{n
}\left(2-\frac{1}{n}\right)=0$. Thus, the subspace $\cL_+$ {\it cannot be
uniformly positive.}

The property of maximality of $\cL_+$ with respect to $(\cdot,\cdot)_{\cP\cT}$
follows from the theory of Krein spaces \cite[Chap. 1]{R3} and the formula
(\ref{e21}). Similar arguments show that $\cL_-$ is a maximally negative
subspace with respect to $(\cdot,\cdot)_{\cP\cT}$. The obtained direct sum
(\ref{e8}) of $\cL_\pm$ is densely defined in $L_2(\mathbb{R})$ and the
corresponding operator $\cC$ is determined by (\ref{e31}). Elementary
calculations using (\ref{e20}) and (\ref{e31}) lead to the conclusion that
\begin{equation}\fl\label{e22}
\cC\phi=\sum_{n=1}^{\infty}\frac{1}{1-\alpha_n^2}\left([(1+\alpha_n^2)c_n^++2i
\alpha_n{c_n^-}]\gamma_n^++[-(1+\alpha_n^2)c_n^-+2i\alpha_n{c_n^+}]\gamma_n^-
\right).
\end{equation}

Let us fix the function $\chi{\in}L_2(\mathbb{R})$,
\begin{equation}\label{e23}
\chi=\sum_{n=1}^{\infty}\frac{1}{n^\delta}(\gamma_n^++\gamma_n^-),\qquad
\frac{1}{2}<\delta\leq\frac{3}{2}
\end{equation}
and set
\begin{equation}\label{e24}\fl
M_{even}=\{\gamma^+\in{L_2^{even}}: (\gamma^+,\chi)=0\},\qquad M_{odd}=\{
\gamma^-\in{L_2^{odd}} : (\gamma^-,\chi)=0\}.
\end{equation}
The functions $\gamma^+=\sum_{n=1}^{\infty}c_n^+\gamma_n^+\in{M_{even}}$ and
$\gamma^-=\sum_{n=1}^{\infty}c_n^-\gamma_n^-\in{M_{odd}}$ can be also
characterized by the condition
\begin{equation}\label{e25}
\sum_{n=1}^{\infty}\frac{c_n^+}{n^\delta}=\sum_{n=1}^{\infty}\frac{c_n^-}
{n^\delta}=0.
\end{equation}

The subspaces
\begin{equation}\fl\label{e26}
\cL_+'=\{f^+=\gamma^++T\gamma^+:\gamma^+\in{M_{even}}\}, \quad \cL_-'=\{f^-=
\gamma^-+T\gamma^-:\gamma^-\in{M_{odd}}\}
\end{equation}
are proper subspaces of $\cL_+$ and $\cL_-$, respectively [since $M_{even}
\subset{L_2^{even}}$ and $M_{odd}\subset{L_2^{odd}}$]. Therefore, the subspaces
$\cL_\pm'$ {\it lose the property of maximality.} We will show that the direct
sum (\ref{e19}) of $\cL_\pm'$ is {\it densely defined} in $L_2(\mathbb{R})$. To
this end, we suppose that a function $y=\sum_{n=1}^{\infty}(y_n^+\gamma_n^++
y_n^-\gamma_n^-)$ is orthogonal to (\ref{e19}). It follows from (\ref{e20}) and
(\ref{e26}) that the condition $y\perp\cL_\pm'$ is equivalent to the relations
\begin{equation*}
\sum_{n=1}^{\infty}(y_n^+-i\alpha_ny_n^-)c_n^+=\sum_{n=1}^{\infty}(y_n^-+i
\alpha_ny_n^+)c_n^-=0,
\end{equation*}
where $\{c_n^\pm\}$ are arbitrary elements of the Hilbert space $l_2$ that also
satisfy (\ref{e25}); that is, $\{c_n^\pm\}$ are orthogonal to the element
$\{1/n^\delta\}$ in $l_2$. This means that
\begin{equation}\label{e27}
y_n^+-i\alpha_ny_n^-=\frac{k_1}{n^\delta}, \qquad y_n^-+i\alpha_ny_n^+=
\frac{k_2}{n^\delta},
\end{equation}
where the constants $k_j$ do not depend on $n$. It follows from (\ref{e27}) that
\begin{equation}\label{e28}
\frac{1}{n}\left(2-\frac{1}{n}\right)y_n^+=(1-\alpha_n^2)y_n^+=(k_1+i\alpha_n{
k_2})\frac{1}{n^\delta}.
\end{equation}

Since the sequence $\{y_n^+\}$ belongs to the Hilbert space $l_2$ and $\delta
\leq\frac{3}{2}$, the relation (\ref{e28}) is possible for $k_1=k_2=0$ only.
Then $y=0$ and the direct sum (\ref{e19}) is densely defined in $L_2(\mathbb{
R})$.

The next step involves the interpretation of $\cL_\pm'$ as the closure
of linear spans of $\cPT$ orthogonal functions $\{f_n^{\pm}\}$. These functions
can be determined in different ways. A `constructive' approach uses (\ref{e20})
to establish that
\begin{equation*}
(I-T^2)\phi=\sum_{n=1}^{\infty}(1-\alpha_n^2)(c_n^+\gamma_n^++c_n^-\gamma_n^-)
\qquad[\phi\in{L_2(\mathbb{R})}].
\end{equation*}
Since $\lim_{n\to\infty}(1-\alpha_n^2)={0}$, the operator $I-T^2$ is compact
selfadjoint in $L_2(\mathbb{R})$ \cite{R14}. Therefore its restrictions
$P_{even}(I-T^2)P_{even}$ and $P_{odd}(I-T^2)P_{odd}$ onto\footnote{$P_{even}$
and $P_{odd}$ are orthogonal projections onto $M_{even}$ and $M_{odd}$ in $L_2(
\mathbb{R})$.} $M_{even}$ and $M_{odd}$ are compact self-adjoint operators in
the Hilbert spaces $M_{even}$ and $M_{odd}$, respectively. These operators have
complete sets of orthonormal eigenfunctions $\{{\gamma_n^+}'\}$ \ [$\{{
\gamma_n^-}'\}$]\ in $M_{even}$ \ [$M_{odd}$], which corresponds to simple
eigenvalues [because $I-T^2$ has the simple eigenvalues $1-\alpha_n^2$].

Denote
\begin{equation*}
f_n^+={\gamma_n^+}'+T{\gamma_n^+}',\qquad f_n^-={\gamma_n^-}'+T{\gamma_n^- }'.
\end{equation*}
The functions $\{f_n^{\pm}\}$ are $\cPT$ orthogonal. Indeed, $(f_n^+,
f_m^-)_{\cPT}=0$ because $f_n^{\pm}\in\cL_\pm'$ by the construction and thus
the subspaces $\cL_\pm'$ are $\cPT$ orthogonal. Furthermore,
$$(f_n^+, f_m^+)_{\cPT}=({\gamma_n^+}'-T{\gamma_n^+}', {\gamma_m^+}'+T{
\gamma_m^+}')=((I-T^2){\gamma_n^+}', {\gamma_m^+}')=\mu_n({\gamma_n^+}',
{\gamma_m^+}')=\mu_n\delta_{nm},$$
where $\mu_n$ is the eigenvalue of $P_{even}(I-T^2)P_{even}$, which corresponds
to the eigenfunction ${\gamma_n^+}'\in{M_{even}}$. Similarly,
\begin{eqnarray*}
(f_n^-, f_m^-)_{\cPT}&=&(-{\gamma_n^-}'+T{\gamma_n^-}', {\gamma_m^-}'+T{
\gamma_m^-}')\\ &=&-((I-T^2){\gamma_n^-}', {\gamma_m^-}')=-\widetilde{\mu}_n({
\gamma_n^-}', {\gamma_m^-}')=-\widetilde{\mu}_n\delta_{nm},
\end{eqnarray*}
where $\widetilde{\mu}_n$ is the eigenvalue of $P_{odd}(I-T^2)P_{odd}$, which
corresponds to the eigenfunction ${\gamma_n^-}'\in{M_{odd}}$. Hence, the
functions $\{f_n^{\pm}\}$ are $\cPT$ orthogonal.

Assume that a function $f^+\in\cL_+'$ is orthogonal to $\mbox{span}\{f_n^{+}\}$.
Then $f^+=\gamma^++T\gamma^+$ [due to (\ref{e26})] and
\begin{eqnarray*}
0&=&(f^+, f_n^{+})=(\gamma^++T\gamma^+,{\gamma_n^+}'+T{\gamma_n^+}')\\
&=&(\gamma^+,{\gamma_n^+}')+(T\gamma^+,T{\gamma_n^+}')=(\gamma^+, (I+T^2){
\gamma_n^+}')=(2-\mu_n)(\gamma^+,{\gamma_n^+}'),
\end{eqnarray*}
where $2-\mu_n\not=0$. This means that the function $\gamma^+\in{M_{even}}$ is
orthogonal to the basis $\{{\gamma_n^+}'\}$ of $M_{even}$. Hence, $\gamma^+=0$
and the closure of $\mbox{span}\{f_n^{+}\}$ coincides with $\cL_+'$. Similar
arguments show that the closure of $\mbox{span}\{f_n^{-}\}$ coincides with
$\cL_-'$.

\emph{We interpret $\{f_n^{\pm}\}$ as eigenfunctions of a Hamiltonian $H$}.
Since the direct sum (\ref{e19}) is a dense subset in $L_2(\mathbb{R})$, the
same property holds true for $\mbox{span}\{f_n^{\pm}\}$. Hence $\{f_n^{\pm}\}$
is \emph{a complete system of eigenfunctions} of $H$ in $L_2(\mathbb{R})$.

\emph{The Hamiltonian $H$ is $\cPT$-symmetric} because its eigenfunctions
$f_n^{\pm}$ are also eigenfunctions of $\cPT$. This property of $f_n^{\pm}$
follows from the construction above. Indeed, the operator $T$ commutes with
$\cPT$ [since $\{T,\cP\}=\{T, \cT\}=0$], and the subspaces $\cL_\pm$ in
(\ref{e21}) are $\cPT$ invariant. The same holds true for the subspaces
$M_{even}$ and $M_{odd}$. [This follows from (\ref{e24}) and the definition of
$\chi$.] The orthogonal projections $P_{even}$ and $P_{odd}$ commute with
$\cPT$. Hence, the operators $P_{even}(I-T^2)P_{even}$ and $P_{odd}(I-T^2)
P_{odd}$ commute with $\cPT$ and their eigenfunctions $\{{\gamma_n^\pm}'\}$ are
eigenfunctions of $\cPT$ (because they have simple eigenvalues!). Thus, $f_n^+=
{\gamma_n^+}'+T{\gamma_n^+}'$ and $f_n^-={\gamma_n^-}'+T{\gamma_n^-}'$ are
eigenfunctions of $\cPT$.

The complete set of eigenfunctions $\{f_n^{\pm}\}$ of $H$ determines the direct
sum (\ref{e19}) [because $\cL_+'$ and $\cL_-'$ are the closures of $\mbox{span}
\{f_n^{+}\}$ and $\mbox{span}\{f_n^{-}\}$, respectively]. However, as shown
above, the subspaces $\cL_\pm'$ are not maximal with respect to the $\cPT$
inner product. Thus, (\ref{e19}) {\it cannot define an operator $\cC$ with
properties (\ref{e5}), (\ref{e6})} in $L_2(\mathbb{R})$. Let us explain this
point. To this end we denote by $\cC'$ an operator with the domain of
definition $\cD(\cC')=\cL_+'[\dot{+}]\cL_-'$ that acts as the identity (minus
identity) operator on $\cL_+'$ ($\cL_+'$). The subspaces $\cL_\pm'$ are $\cPT$
invariant because $f_n^{\pm}$ are eigenfunctions of $\cPT$.

Thus, the operator $\cC'$ satisfies the conditions (\ref{e5}):
$${\cC'}^2f=f, \quad \cC'\cPT{f}=\cPT\cC'{f} \quad [f\in\cD(\cC')] \quad
\mbox{and}\quad \cC'{Hg}=H\cC'{g} \quad [g\in\cD(H)],$$
where $\cD(H)$ is contained in $\cD(\cC')$ by the definition of $H$. However,
$\cC'$ {\it cannot satisfy (\ref{e6})} and, as a result, $\cC'$ {\it cannot be
presented as (\ref{e7})}. Indeed, the assumption that $\cC'=e^\cQ\cP$, where
$e^\cQ$ is a {\it Hermitian operator}, leads to the conclusion\footnote{This
conclusion is established by repeating the proof of statement I in the
Appendix.} that the subspaces $\cL_\pm'$ have the form
$$\cL_+'=\{f^+=\gamma^++T\gamma^+:\gamma^+\in{L_2^{even}}\},\qquad\cL_-'=\{f^-=
\gamma^-+T\gamma^-:\gamma^-\in{L_2^{odd}}\},$$
where $T=\tanh{\frac{{\cQ}}{2}}$ is a Hermitian strong contraction {\it defined
on} $L_2(\mathbb{R})$. This contradicts the original presentation (\ref{e26})
of $\cL_\pm'$, where $T$ is defined on $M_{even}{\oplus}M_{odd}\subset{L_2(
\mathbb{R})}$. In particular, the operator $T$ in (\ref{e26}) is not defined on
$\chi\in{L_2(\mathbb{R})}$.

Therefore, the metric operator $e^\cQ=\cC'\cP$ determined by the eigenfunctions
$\{f_n^{\pm}\}$ of $H$ {\it cannot be Hermitian}. This obstacle can be removed
if we extend $\cL_\pm'$ to maximal positive/negative subspaces $\cL_\pm$ with
respect to the $\cPT$ inner product. In this case the operator $\cC'$ is
extended to an operator $\cC$ with properties (\ref{e5}), (\ref{e6}) [that is,
we guarantee the Hermiticity of the metric operator $\cC\cP$ by extending the
domain $\cD(\cC')$].

Additional calculations with the use of Theorem 4 in \cite{R17} show that if
the parameter $\delta$ in (\ref{e23}) satisfies the condition $1<\delta\leq
\frac{3}{2}$, then the pair $\cL_+'[\dot{+}]\cL_-'$ can be extended to
\emph{different} pairs of maximal positive/negative subspaces $\cL_+[\dot{+}]
\cL_-$. In this case we have different extensions $\cC\supset\cC'$ that satisfy
(\ref{e5}) and (\ref{e6}). These extensions are different unbounded $\cC$
operators for the $\cPT$-symmetric Hamiltonian $H$ defined above. One of the
possible extensions is the operator $\cC$ defined by (\ref{e22}). The
corresponding pair of maximal subspaces $\cL_\pm$ is determined by (\ref{e21}).

On the other hand, if $\frac{1}{2}<\delta\leq{1}$, then the extension of
$\cL_+'[\dot{+}]\cL_-'$ to a maximal pair $\cL_+[\dot{+}]\cL_-$ is {\it
unique}. The subspaces $\cL_\pm$ are determined by (\ref{e21}) and the formula
(\ref{e22}) provides the unique extension $\cC\supset\cC'$ with properties
(\ref{e5}) and (\ref{e6}). In this case the $\cPT$-symmetric Hamiltonian $H$
has the unique unbounded operator $\cC$.

\section{Conclusions}
\label{s6}

This paper shows that if the $\cC$ operator for a $\cPT$-symmetric Hamiltonian
with simple eigenvalues is nonunique, then it is {\it unbounded}. (A bounded
$\cC$ operator can be constructed for finite-matrix Hamiltonians and for
Hamiltonians generated by differential expressions with $\cPT$-symmetric point
interactions.) As a consequence, the mapping between a conventionally Hermitian
Hamiltonian and a $\cPT$-symmetric Hamiltonian having real eigenvalues is
unbounded. Thus, while there is a formal mapping between the Hilbert spaces of
the two theories, the mapping does not map all of the vectors in the domain of
one Hamiltonian into the domain of the other Hamiltonian. Consequently, even if
the conventionally Hermitian Hamiltonian and the $\cPT$-symmetric Hamiltonian
are isospectral, they are mathematically inequivalent theories. Thus, at a
fundamental mathematical level a $\cPT$-symmetric Hamiltonian describes a
theory that is new. It is an open question whether it is possible to design
physical experiments that can detect the differences between these two
theories.

\section{Appendix}
To show the equivalence of statements I and II, we consider the $\cPT$
orthogonal decomposition of $L_2(\mathbb{R})$ onto its even $L_2^{even}$ and
odd $L_2^{odd}$ subspaces
\begin{equation}\label{e29}
L_2(\mathbb{R})=L_2^{even}[\dot{+}]L_2^{odd},
\end{equation}
which are, respectively, maximal positive and maximal negative with respect to
$(\cdot,\cdot)_{\cPT}$. The subspaces $\cL_+$ and $\cL_-$ in (\ref{e8}) are
also maximal positive and maximal negative and their `deviation' from
$L_2^{even}$ and $L_2^{odd}$ is described by a Hermitian strong contraction
$T$, which anticommutes with $\cP$ \cite{R15}. To be precise,
\begin{equation}\fl\label{e29b}
\cL_+=\{f^+=\gamma^++T\gamma^+\ :\ \gamma^+\in{L_2^{even}}\},\quad\cL_-=\{f^-=
\gamma^-+T\gamma^-\ :\ \gamma^-\in{L_2^{odd}}\}.
\end{equation}

Denote by $P_{\pm}$ the projection operators onto $\cL_\pm$ in $L_2(\mathbb{R}
)$. The operators $P_{\pm}$ are defined on the linear set $\cD=\cL_+[\dot{+}]
\cL_-$ and
\begin{equation*}
P_+(f^++f^-)=f^+, \qquad P_-(f^++f^-)=f^- \qquad [f^\pm\in\cL_\pm].
\end{equation*}
These operators $P_{\pm}:\cD\to\cL_\pm$ can also be determined by the formulas
\begin{equation}\label{e30}
P_+=(I-T)^{-1}(P_{even}-TP_{odd}), \qquad P_-=(I-T)^{-1}(P_{odd}-TP_{even}),
\end{equation}
where $P_{even}=\frac{1}{2}(I+\cP)$ and $P_{odd}=\frac{1}{2}(I-\cP)$ are
projections on $L_2^{even}$ and $L_2^{odd}$.

Let us prove (\ref{e30}). First, we note that $P_{even}T=TP_{odd}$ since $\{T,
\cP\}=0$. Then, for any function $\phi=\gamma^++\gamma^-$ \ [$\gamma^+{\in}
L_2^{even}$, $\gamma^-\in{L_2^{odd}}$] from $L_2(\mathbb{R})$,
\begin{eqnarray*}
P_+(I+T)\phi=(I+T)\gamma^+=(I+T)P_{even}\phi=(I-T)^{-1}(I-T^2)P_{even}\phi=& &
\\ (I-T)^{-1}P_{even}(I-T^2)\phi=(I-T)^{-1}(P_{even}-TP_{odd})(I+T)\phi, & &
\end{eqnarray*}
which establishes the first formula in (\ref{e30}) because $(I+T)L_2(\mathbb{R}
)=\cD$ due to (\ref{e29}) and (\ref{e29b}). The second formula is proved by
similar arguments.

{\it Proof of statement II.} Let the subspaces $\cL_\pm$ in (\ref{e8}) be
$\cPT$ invariant and let $\cC$ act as the $\pm$identity operator on $\cL_\pm$.
Then, using (\ref{e30}), we obtain
\begin{equation}\fl\label{e31}
\cC=P_+-P_-=(I-T)^{-1}(P_{even}-TP_{odd}-P_{odd}+TP_{even})=(I-T)^{-1}(I+T)\cP.
\end{equation}

The spectrum of $T$ is contained in the segment $[-1,1]$ and $\pm{1}$ cannot be
eigenvalues of $T$ because $T$ is a Hermitian strong contraction. In such a case
$$(I-T)^{-1}(I+T)=e^\cQ,~~\mbox{where}~~\cQ=s(T)~~\mbox{and}~~s(\lambda)=\ln
\frac{1+\lambda}{1-\lambda},$$
is a Hermitian operator in $L_2(\mathbb{R})$.

It follows from (\ref{e29b}) that the $\cPT$ invariance of $\cL_\pm$ is
equivalent to the commutation relation $[T,\cPT]=0$. Then $[e^\cQ,\cPT]=0$ and
hence, $[\cQ,\cPT]=0$.

On the other hand, the condition $\{T,\cP\}=0$ implies that the spectral
function $E_\lambda$ of $T$ satisfies the relation ${\cP}E_\Delta=E_{-\Delta}
\cP$ for any interval $\Delta$ of $\mathbb{R}$ \cite{R16}. Using this relation
and the fact that $s(\lambda)=\ln\frac{1+\lambda}{1-\lambda}$ is an odd
function on $[-1,1]$, we obtain
$$\cP\cQ=\cP\int_{[-1,1]}s(\lambda)dE_\lambda=\int_{[-1,1]}s(\lambda)dE_{-
\lambda}\cP=-\int_{[-1,1]}s(-\lambda)dE_{-\lambda}\cP=-\cQ\cP.$$
Hence, $\{\cQ,\cP\}=0$. Combining this with $[\cQ,\cPT]=0$ we conclude that $\{
\cQ,\cT\}=0$. Thus $\cC$ is determined by (\ref{e7}), where $\{\cQ,\cP\}=\{\cQ,
\cT\}=0$.

{\it Proof of statement I.} Let $\cC$ be determined by (\ref{e7}). In this
representation, $e^\cQ$ is a positive Hermitian operator. This means that the
operator
$$T=(e^{\cQ}-I)(e^{\cQ}+I)^{-1}=\frac{e^{{\cQ}/2}-e^{-{\cQ}/2}}{2}\left(\frac{
e^{{\cQ}/2}+e^{-{\cQ}/2}}{2}\right)^{-1}=\frac{\sinh({\cQ}/2)}{\cosh({\cQ}/2)}=
\tanh{\frac{{\cQ}}{2}}$$
is a Hermitian strong contraction defined on $L_2(\mathbb{R})$. Moreover, $T$
anticommutes with $\cP$ and $\cT$ since $\{\cQ,\cP\}=\{\cQ,\cT\}=0$.
With the help of $T$ and (\ref{e29b}), we determine the maximal positive
(negative) subspaces $\cL_\pm$ in the direct sum (\ref{e8}). The subspaces
$\cL_\pm$ are $\cPT$ invariant. The operator $\cC$ corresponding to (\ref{e8})
is determined by (\ref{e31}); that is,
$$\cC=(I-T)^{-1}(I+T)\cP=\left(I-\tanh\frac{\cQ}{2}\right)^{-1}\left(I+\tanh
\frac{\cQ}{2}\right)\cP=e^{\cQ}\cP.$$
Therefore, the domain of definition $\cD(\cC)$ is determined by (\ref{e8}) and
$\cC=e^{\cQ}\cP$ acts as the $\pm$ identity operator on elements of $\cL_\pm$.

Using (\ref{e31}) again we establish (\ref{e10}):
$$\frac{1}{2}(I\ \pm\ \cC)\cD(\cC)=\frac{1}{2}(I \ \pm(P_+-P_-))\cD(\cC)=P_\pm
\cD(\cC)=\cL_\pm.$$

\section*{Acknowledgments} CMB is supported by grants from the U.S. Department
of Energy and the U.K. Leverhulme Foundation. SK thanks the Fulbright
Foundation for research support and Washington University in St.~Louis for its
warm hospitality.

\end{document}